\documentclass[a4paper,11pt]{article}

\pdfoutput=1 % if your are submitting a pdflatex (i.e. if you have
\usepackage{enumerate} 
\usepackage{afterpage} 
\usepackage{amsmath} 
\usepackage{ulem}
\usepackage{amssymb} 
\usepackage{jcappub} % for details on the use of the package, please
\usepackage{soul}
\usepackage{lineno}
\usepackage[T1]{fontenc} % if needed
\usepackage[usenames,dvipsnames]{xcolor}

\usepackage[colorinlistoftodos]{todonotes} % todo notes
\usepackage{textcomp} % for \textmu

\title{\boldmath Search for magnetic inelastic dark matter with XENON100}

\author[]{The XENON collaboration:\\}
\author[a]{E.~Aprile,}%\columbia
\author[b]{J.~Aalbers,}%\nikhef
\author[c,d]{F.~Agostini,}%\lngs\bologna
\author[e]{M.~Alfonsi,}%\mainz
\author[f]{F.~D.~Amaro,}%\coimbra
\author[a]{M.~Anthony,}%\columbia
\author[g]{F.~Arneodo,}%\nyuad
\author[h]{P.~Barrow,}%\zurich
\author[h]{L.~Baudis,}%\zurich
% ...
\affiliation[a]{Physics Department, Columbia University, New York, NY 10027, USA}
\affiliation[b]{Nikhef and the University of Amsterdam, Science Park, 1098XG Amsterdam, Netherlands}
\affiliation[c]{INFN-Laboratori Nazionali del Gran Sasso and Gran Sasso Science Institute, 67100 L'Aquila, Italy}
\affiliation[d]{Department of Physics and Astrophysics, University of Bologna and INFN-Bologna, 40126 Bologna, Italy}
\affiliation[e]{Institut f\"ur Physik \& Exzellenzcluster PRISMA, Johannes Gutenberg-Universit\"at Mainz, 55099 Mainz, Germany}
\affiliation[f]{LIBPhys, Department of Physics, University of Coimbra, 3004-516 Coimbra, Portugal}
\affiliation[g]{New York University Abu Dhabi, Abu Dhabi, United Arab Emirates}
\affiliation[h]{Physik-Institut, University of Zurich, 8057 Zurich, Switzerland}
% ...

\emailAdd{lukas.buetikofer@lhep.unibe.ch}
\emailAdd{xenon@lngs.infn.it}

\abstract{We present the first search for dark matter-induced delayed coincidence signals in a dual-phase xenon time projection chamber, using the 224.6\,live days of the XENON100 science run~II. This very distinct signature is predicted in the framework of magnetic inelastic dark matter which has been proposed to reconcile the modulation signal reported by the DAMA/LIBRA collaboration with the null results from other direct detection experiments. No candidate event has been found in the region of interest and upper limits on the WIMP's magnetic dipole moment 
are derived. The scenarios proposed to explain the DAMA/LIBRA modulation signal by magnetic inelastic dark matter interactions of WIMPs with masses of 58.0\,GeV/c$^2$ and 122.7\,GeV/c$^2$ are excluded at 3.3\,$\sigma$ and 9.3\,$\sigma$, respectively.}

\begin{document}
\maketitle
\flushbottom

\section{\label{sec:intro}Introduction}
The existence of dark matter in the Universe has been inferred indirectly from a large number of astrophysical and cosmological observations at all length scales~\cite{Bergstrom:DMreview}. A plethora of experiments based on different techniques aim at the direct detection of dark matter in sensitive underground detectors~\cite{Schumann:DM,Baudis:DMdetection, Marrodan:DMexperiments}. However, the limits on the dark matter-ordinary matter scattering cross section derived by these experiments are in strong conflict with the long-standing dark matter detection claim by DAMA/LIBRA~\cite{Bernabei:2015}, especially if their 9.3\,$\sigma$ modulation signal is interpreted within the usual framework of Weakly Interacting Massive Particles (WIMPs) \cite{Savage:2008}. Several alternatives to the classical WIMP scenario have been proposed in order to reconcile the null results of other experiments with  DAMA/LIBRA. One of these models is {\it magnetic inelastic dark matter} (MiDM) proposed by Chang~et~al.~\cite{Chang:2010}.

Similar to inelastic dark matter (iDM)~\cite{TuckerWeiner:iDM}, MiDM is based on the assumption that there is an excited WIMP state~$\chi^{\star}$ with a corresponding mass splitting~$\delta$. Furthermore, inelastic scattering of the WIMP against the nucleus is allowed, while elastic scattering is highly suppressed or forbidden. In addition, MiDM assumes that WIMPs have a non-zero magnetic dipole moment~$\mu_\chi$. The finite mass splitting~$\delta$ requires a minimal velocity 
\begin{equation}
v_\text{min} = \frac{1}{\sqrt{2M_N E_R}}\left(\frac{M_N E_R}{\overline{\mu}} + \delta \right)
\end{equation}
for a WIMP to scatter off a nucleus. $M_N$ is the mass of the target nucleus, $E_R$ the nuclear recoil energy and $\overline{\mu}$ is the reduced mass of the WIMP-nucleus system. This restriction favors heavy targets, such as iodine used in DAMA/LIBRA  ($A$\,=\,126.9) or xenon ($A$\,=\,131.3), since the WIMP spectrum gets shifted to higher energies.

Due to the WIMP magnetic moment MiDM features dipole-dipole (DD) as well as dipole-charge (DZ) interactions between the WIMP and the target nucleus. These interactions favor iodine thanks to its large nuclear magnetic moment ($\mu_\mathrm{I}=2.8\mu_\mathrm{nuc}$) compared to most targets typically used by other dark matter experiments. Taking into account the high mass number and the large magnetic moment of iodine, MiDM opens up new parameter space for the DAMA/LIBRA modulation signal which is not in conflict with other null results~\cite{Chang:2010, Barello:2014}.

MiDM interactions lead to two different signatures that can be employed for an essentially background free experimental search. The first is a single-scatter nuclear recoil signal from the WIMP-nucleus interaction, however, with a higher mean recoil energy~$E_R$ compared to the ``standard'' spin-independent interaction~\cite{XENON100:iDM}. The second signature is a distinct feature of the MiDM model: the excited WIMP de-excites with a lifetime~$\tau=\pi/(\delta^3\mu_\chi^2) \approx \mathcal{O}$($\mu$s) (for the values of $\delta$ and $\mu_\chi$ considered in this analysis). During this period, the WIMP propagates a distance of $\mathcal{O}$(m) given the mean velocity of the Sun with respect to the WIMP halo. The de-excitation leads to the emission of a $\mathcal{O}$(100\,keV) photon which will interact with the target as well, inducing an electronic recoil signal. This unique combination of a low-energy nuclear recoil followed by a significantly larger electronic recoil provides the means for the first search for dark matter-induced interactions in double-scatter signatures.

For the analysis presented here, we use data from the science run~II of the XENON100 dark matter experiment, previously used for various analyses~\cite{xenon100:225live_days,XENON100:Leptophilic,XENON100:Modulation,XENON100:Axion}. The data was acquired between February 28, 2011 and March 31, 2012 comprising a total live time of 224.6~days. XENON100, a liquid xenon time projection chamber (LXe TPC) described in detail in \cite{XENON100:instrument}, is located at the Laboratori Nazionali del Gran Sasso (LNGS) of INFN in Italy. The TPC is instrumented with two arrays of photomultipliers (PMTs, Hamamatsu R8520), one below the 62\,kg LXe target in the cryogenic liquid and one above in the xenon gas phase. A particle interaction inside the TPC leads to a prompt scintillation signal (S1) and liberates free ionization electrons, that are drifted towards the liquid-gas interface by an electric field of  0.53\,kV/cm. A stronger electric field ($\sim$12\,kV/cm) extracts them into the gas phase, where they create a secondary scintillation signal (S2), which is proportional to the ionization charge~\cite{dualphase}. The interaction vertex can be spatially reconstructed using the time separation of the two signals and the S2-signal spatial distribution on the top PMT array.  The ratio of scintillation light and ionization charge signal depends on the interacting particle. This allows the discrimination of $\gamma$ and $\beta$ backgrounds, which produce electronic recoils (ER), from nuclear recoils (NR) that are expected from WIMP interactions.

\begin{figure}[b!]
\centering
\begin{minipage}[t]{.40\textwidth}
  \centering
  \includegraphics[width=1.\linewidth]{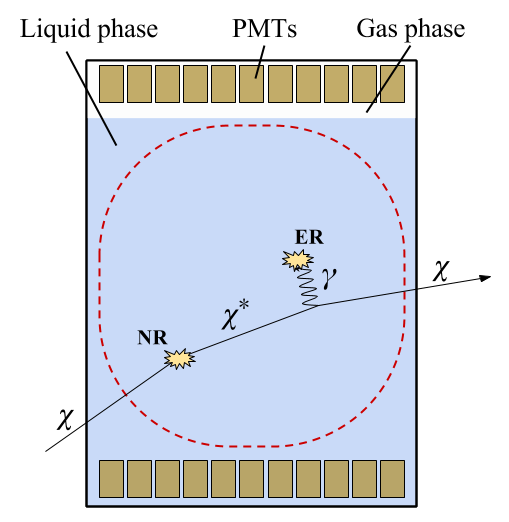}
\end{minipage}%
\begin{minipage}[t]{.55\textwidth}
  \centering
  \includegraphics[width=1.05\linewidth]{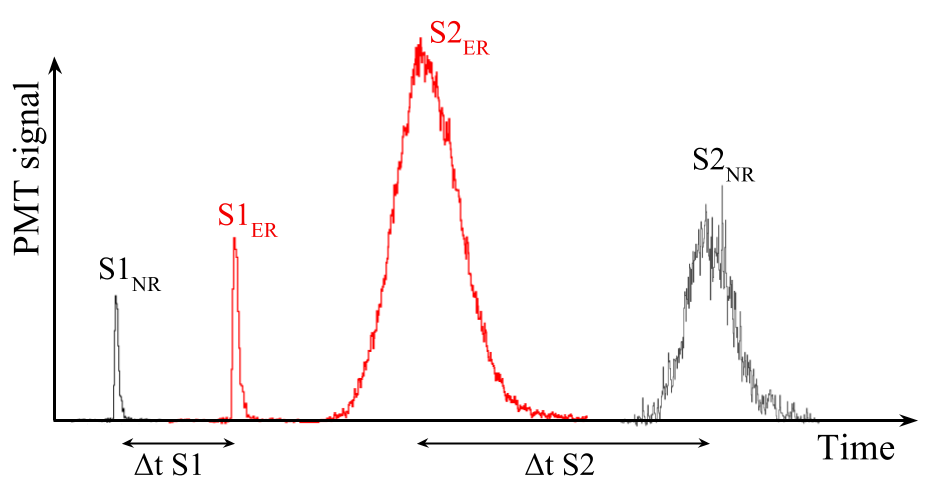}
\end{minipage}
\caption{\textbf{(Left)} The expected signature from the interaction of magnetic inelastic dark matter consists of a primary WIMP-nucleon scattering (NR signal) and the subsequent decay of the excited WIMP, leading to a $\gamma$-emission (ER signal in TPC). In the analysis, both interactions have to happen within the 48\,kg fiducial volume illustrated by the dashed line. \textbf{(Right)} Illustration of the expected PMT waveform corresponding to the interaction shown on the left. Peaks corresponding to the NR (ER) interaction are shown in black (red). The narrow peaks on the left are S1~signals, the wider ones on the right S2 signals. The first~S1 peak always corresponds to the NR interaction.}
\label{fig:double_interaction}
\end{figure}

The size of the cylindrical XENON100 TPC ($\sim$30\,cm diameter and height) allows a first-ever search for the distinct MiDM signature of a primary nuclear recoil followed by the photon emitted by the WIMP de-excitation, as illustrated in Figure~\ref{fig:double_interaction}. Thanks to the low background expectation for this event topology, a large fiducial target of 48 kg can be used for the analysis. This is 40\% higher than employed for previous searches using the same dataset, thereby increasing the detection efficiency for the MiDM-interaction. In order to compare our result to the DAMA/LIBRA signal we focus on two WIMP masses, 58.0\,GeV/c$^2$ and 122.7\,GeV/c$^2$, which correspond to the best-fit results from Ref.~\cite{Barello:2014} to explain the modulation within the MiDM model. The first mass corresponds to an iodine quenching factor of $Q_\text{I}=0.09$~\cite{Bernabei:1996}, the second one to a more recently measured value of $Q_\text{I}=0.04$~\cite{Collar:Q_I}. Since the inelastic kinematics favors heavy targets, only scattering off the iodine nuclei in~NaI is considered.
%\cite{Barello:2014}.

%%%%%%%%%%%%%%%%%%%%%%%%%%%%%%%%%%%%%%%%%%%%%%%%%%%%%%%%%%%%%%%%%%%%%%%%%%%%%%
\section{\label{sec:event_rate}Expected event rate} 

The differential event rate is given by
\begin{equation}
\label{eq:diff_rate}
	\frac{dR}{dE_R} = \frac{\rho_0}{m_N \ m_\chi} \int_{v_\mathrm{min}}^{v_\mathrm{max}} vf(\mathbf{v})\frac{\text{d}\sigma}{\text{d}E_R}\text{d}^3v \textnormal{,}
\end{equation} 
with the local dark matter density $\rho_0=0.3$\, GeV/cm$^3$~\cite{Green:DMdensity} and the Maxwell-Boltzmann velocity distribution $f(\mathbf{v})$ from~\cite{Barello:2014} (local circular velocity $v_0=220\,\mathrm{km/s}$, Galactic escape velocity $v_\mathrm{esc}= 550\,\mathrm{km/s}$). The differential cross-section for MiDM-nucleus scattering d$\sigma$/d$E_R$ is a sum of two parts, the dipole-dipole (DD):
\begin{equation}
\label{eq:DD_cross_section}
	\frac{\text{d}\sigma_{DD}}{\text{d}E_{R}} = \frac{16\pi \alpha^2m_N}{v^2}\left(\frac{\mu_{N}}{e}\right)^2\left(\frac{\mu_{\chi}}{e}\right)^2
	\left(\frac{S_\chi\!+\!1}{3S_\chi}\right) \left(\frac{S_N\!+\!1}{3S_N} \right)F^2_D(E_{R})
\end{equation}
and the dipole-charge (DZ) contribution~\cite{Chang:2010}:
%\small
\begin{equation}
\label{eq::dz}
\frac{\text{d}\sigma_{DZ}}{\text{d}E_{R}} = \frac{4\pi Z^2\alpha^2}{E_{R}}\!\left(\frac{\mu_{\chi}}{e}\right)^2\!\left[1\!-\!\frac{E_{R}}{v^2}\!\left(\frac{1}{2m_N}\!+\!\frac{1}{m_\chi}\right)\!\!-\!\frac{\delta}{v^2}\left(\frac{1}{\overline{\mu}}\!+\!\frac{\delta}{2m_N E_{R}}\right)\!\right]\!\!\left(\frac{S_\chi\!+\!1}{3S_\chi}\right) F^2(E_{R}) \textnormal{.}
\end{equation}
%\normalsize
$\mu_\chi$ and $\mu_N$ are the magnetic dipole moment of the WIMP and of the target nucleus, respectively. $\alpha$ denotes the fine structure constant. Natural xenon contains two isotopes with a non-zero magnetic moment, $^{131}$Xe and $^{129}$Xe with an abundance of 21.2\,\% and 26.4\,\%, and is thus sensitive to DD~interactions. $F^2_D(E_R)$ is the magnetic dipole form-factor from~\cite{Barello:2014} and $F^2(E_R)$ is the Helm form-factor~\cite{Lewin:FF}. Finally, as in Ref.~\cite{Barello:2014}, the spin of the WIMP is assumed to be $S_\chi=1/2$ and $S_N$ denotes the nuclear spin of xenon.

\begin{figure}[b]
  \begin{minipage}[c]{0.5\textwidth}
    \includegraphics[width=0.95\textwidth]{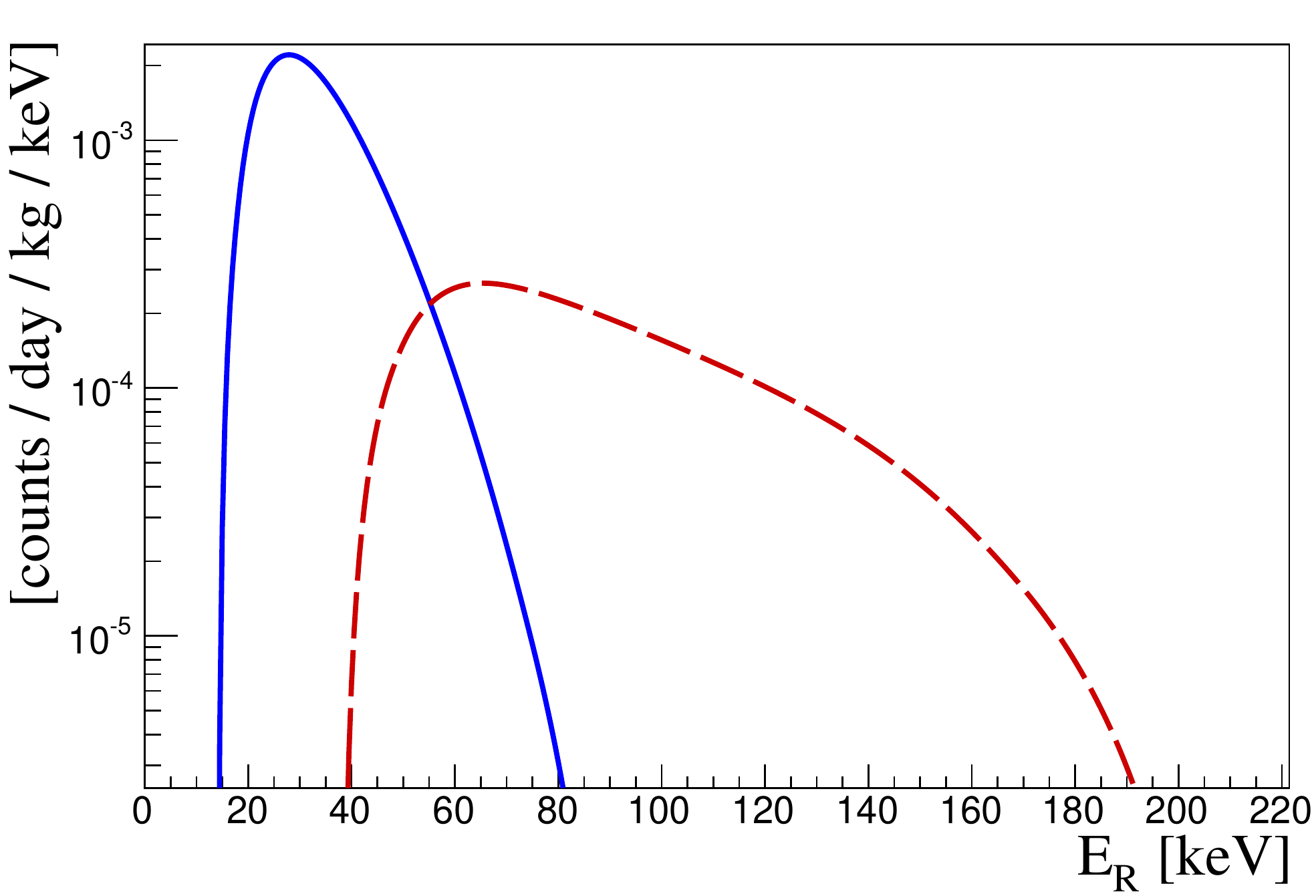}
  \end{minipage}\hfill
  \begin{minipage}[c]{0.5\textwidth}
  \vspace{-1.8cm}
    \caption{Expected nuclear recoil energy spectra in counts per day per~kg and~keV for the two sets of parameters $(m_\chi,\ \mu_\chi,\ \delta)$, corresponding to the benchmark cases~1 (solid blue: $m_\chi=58.0\,\mathrm{GeV/c^2},\ \mu_\chi=0.0019\, \mu_{\mathrm{nuc}},\ \delta=111.7\,\mathrm{keV}$) and~2 (dashed red: $m_\chi=122.7\,\mathrm{GeV/c^2},\ \mu_\chi=0.0056\, \mu_{\mathrm{nuc}},\ \delta=179.3\,\mathrm{keV}$).} 
    \label{fig:energy_spec}
  \end{minipage}
\end{figure}

The three free parameters in the analysis are the WIMP mass~$m_\chi$, magnetic dipole moment~$\mu_\chi$ and mass splitting~$\delta$. The expected energy spectrum for a given set of  parameters $(m_\chi,\ \mu_\chi,\ \delta)$ is calculated as in Ref.~\cite{Barello:2014}, using a modified code originally provided by the authors from~\cite{Fitzpatrick:MathematicaPackage,Anand:MathematicaPackage}.
Figure~\ref{fig:energy_spec} shows the expected nuclear recoil energy spectra for the two benchmark cases corresponding to the DAMA/LIBRA best fit values for the different quenching factors $Q_\text{I}$:
\begin{enumerate}
\item $Q_\text{I}=0.09$: $(m_\chi=58.0\,\mathrm{GeV/c^2},\ \mu_\chi=0.0019\, \mu_{\mathrm{nuc}},\ \delta=111.7\,\mathrm{keV})$
\item $Q_\text{I}=0.04$: $(m_\chi=122.7\,\mathrm{GeV/c^2},\ \mu_\chi=0.0056\, \mu_{\mathrm{nuc}},\ \delta=179.3\,\mathrm{keV})$,
\end{enumerate}
with the nuclear magneton $\mu_{\mathrm{nuc}}$. Both spectra agree with the ones presented in~\cite{Barello:2014} and start well above the XENON100 energy threshold of 6.6\,keV$_\mathrm{nr}$~\cite{xenon100:225live_days}. This analysis is thus not limited by the lower energy threshold.

%%%%%%%%%%%%%%%%%%%%%%%%%%%%%%%%%%%%%%%%%%%%%%%%%%%%%%%%%%%%%%%%%%%%%%%%%%%%%%
\section{\label{sec:analysis}Data analysis}

The MiDM event topology exploited in this analysis is a NR~interaction  followed by an~ER of energy $\delta$, induced by the photon emitted in the WIMP de-exitation. Thus the following three chronological sequences of~S1 and S2~signals can possibly be detected in XENON100:
\begin{enumerate}
\item S1$_\text{NR}$ $\longrightarrow$ S1$_\text{ER}$ $\longrightarrow$ S2$_\text{NR}$ $\longrightarrow$ S2$_\text{ER}$
\item S1$_\text{NR}$ $\longrightarrow$ S1$_\text{ER}$ $\longrightarrow$ S2$_\text{ER}$ $\longrightarrow$ S2$_\text{NR}$
\item S1$_\text{NR}$ $\longrightarrow$ S2$_\text{NR}$ $\longrightarrow$ S1$_\text{ER}$ $\longrightarrow$ S2$_\text{ER}$.
\end{enumerate}
Since only single scatter NR~events are expected in a ``standard'' WIMP analysis~\cite{xenon100:225live_days}, the XENON100 peak finding algorithm does not search for S1~signals after the first large S2~peak. Thus, the second S1 peak (S1$_\text{ER}$) in the third topology will be missed. For this reason, only the first two interaction sequences are considered in this analysis. The example shown in Figure~\ref{fig:double_interaction} would correspond to sequence~2. While the time-order of the S1 peaks determines their interaction type (NR followed by ER), the assignment of the two S2~peaks to the corresponding~S1s is based on their energy.

This very distinct ``delayed coincidence'' event topology of two S1 signals followed by two S2~signals allows the removal of most of the backgrounds. The very abundant double scatter processes from Compton-scattering $\gamma$-radiation or neutrons is a negligible background process for this analysis: the S1~signals of such double scatters are generated almost simultaneously, which is why they are removed once a minimal time separation~$\Delta t$ between the two S1~peaks is required. We thus select events with $\Delta t>$50\,ns, the minimum time difference at which two S1~signals can be separated with basically 100\,\% efficiency by the raw data processor. This efficiency was measured using a sample of artificially generated waveforms with two S1~peaks at variable time separations $\Delta t$. The maximal $\Delta t$ of 2\,$\mu s$ covers the vast majority of all possible tracks of a WIMP inside the target, given its velocity distribution and the detector dimensions. The signal loss due to these requirements on $\Delta t$, as well as the impact of ignoring the third event sequence listed above, is taken into account by the efficiency simulation described in Section~\ref{sec:efficiency}.

The analysis of double scatter waveforms showed that a minimum time difference of $3.5$\,$\mu$s between the two ionization signals is required in order to be able to separate them efficiently. This condition~(i) is applied to the data and the corresponding signal acceptance loss is again taken into account by the efficiency simulation. To obtain a high efficiency for the detection of both S2~peaks we place cuts on (ii) the minimal height and (iii) area of the smaller S2~peak. The overall efficiency to detect two S2~peaks that meet the requirements (ii)+(iii) is $>$94\%. It is determined by inspecting a sample of waveforms of $^{241}$AmBe neutron data for unrecognized S2~peaks. Neutron data is used as the smaller S2~peak is required to be from the NR interaction.  
We additionally apply some of the data quality cuts which were already used for previous analyses~\cite{xenon100:225live_days,xe100:analysis} and which were now adapted for the expected MiDM signal. These are (iv) loose cuts on the S1~size, (v) the rejection of events with a coincident S1~signal in the optically separated LXe volume surrounding the TPC (``veto'') and (vi) the selection of events with exactly two S2~peaks which can be causally related to the S1~peak(s). The latter condition requires that the S1~and S2~signals are detected within the maximal drift time of the TPC. Finally, we (vii) reject interactions in the xenon gas phase based on the number of PMTs which detect the S1~signal simultaneously (coincidence level $N_c$). This quantity is energy-dependent and interactions in the gas have lower $N_c$ than events in the liquid. The combined acceptance of cuts (iv)-(vii) is 99\,\%. Loose energy bounds on the S1 and S2~signals allow searching for nuclear recoils in the range $E_R=(9.7-200)$\,keV$_{\mathrm{nr}}$ and for mass splittings  $\delta = (30-200)$\,keV$_\mathrm{ee}$. Their acceptance is 100\,\% for the two studied benchmark cases. Both energy scales are based on the S1~signal, with the relative scintillation efficiency ${\cal L}_\text{eff}$ as defined in~\cite{xenon100:225live_days} and the electronic recoil scale as used in~\cite{XENON100:Axion}. 

\begin{figure}[b!]
\centering
\begin{minipage}[t]{.5\textwidth}
  \centering
  \includegraphics[width=.95\linewidth]{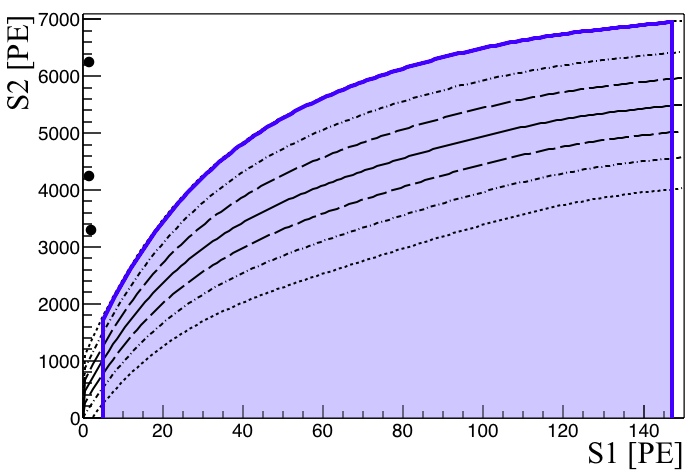}
  \label{fig:ROI}
\end{minipage}%
\begin{minipage}[t]{.5\textwidth}
  \centering
  \includegraphics[width=.95\linewidth]{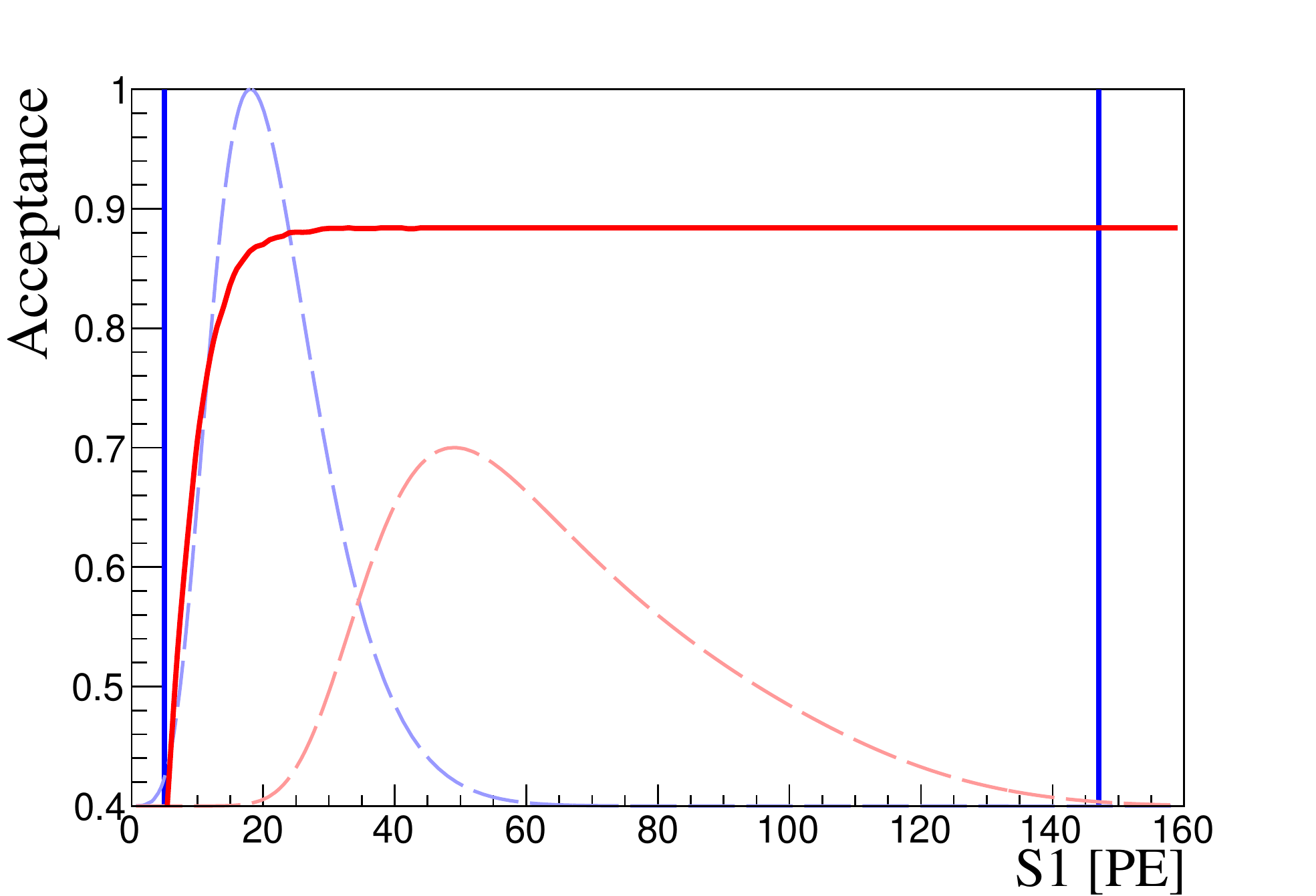}
  \label{fig:combined_acceptance}
\end{minipage}
\caption{\textbf{(Left)} Region of interest (ROI) for the NR signal of a MiDM interaction, defined by the 95\% NR~acceptance cut (top blue line) and the S1-energy interval (left and right blue line). The dashed lines show the $\pm$20\%, $\pm$35\% and $\pm$45\% quantiles of the NR region as defined by $^{241}$AmBe neutron calibration data. No events remain in the ROI after applying all cuts (on all four S1 and S2 peaks). Three remaining events (black dots) appear at very low~S1, outside of the ROI. Their S1$_\text{NR}$ is due to electronic noise, therefore they are no physical events and don't fall into the NR region. \textbf{(Right)} Combined cut acceptance as function of the NR interactions' S1~signal (solid red). The spectra from Figure~\ref{fig:energy_spec} are given for comparison, with the recoil energy being converted to PE taking into account the finite energy resolution~\cite{xe100:analysis}. The vertical lines indicate the analysis energy range of $E_R=(9.7 - 200)$\,keV$_{\mathrm{nr}}$.}
\end{figure}

We additionally require that the scatter associated with the first~S1 has an S2/S1-ratio corresponding to events in the NR region, which has been defined based on $^{241}$AmBe calibration data. By construction, this region includes 95\,\% of the NRs. The region of interest for the NR interaction is defined by this cut and the energy interval, see Fig.~\ref{fig:ROI}~(left). The combined cut acceptance is shown in Fig.~\ref{fig:combined_acceptance}~(right). The low-energy drop is mainly due to the requirements on height and area of the smaller S2~signal. The constant acceptance loss at higher energies is mainly due to the efficiency of the data processor to find two appropriate S2~peaks (94\%) and the requirement that the first interaction falls into the NR region (95\%). Since the expected NR spectra from MiDM are shifted to higher energies, the low-energy acceptance-loss has only little impact on the analysis.

The background expectation for the 48\,kg\,$\times$\,224.6\,days = 10.8\,t\,$\times$\,days exposure studied in this analysis takes into account contributions from the following background sources: (a) pile-up of two individual ER single scatter events; (b) pile-up between a standard double scatter (one~S1 and two~S2 peaks) and a ``lone'' S1 peak without any correlated S2, e.g., from interactions inside a charge insensitive region; (c) delayed $\beta^-\to \gamma$ coincidences from radioactive $^{85}$Kr decays ($T_{1/2} \sim 1\,\mathrm{\mu s}$); and (d) delayed $\beta^-\to \alpha$ coincidences from the Bi-Po decays following $^{220}$Rn (${T_{1/2}= 3}$\,ms) and $^{222}$Rn decays ($T_{1/2}=164.3\,\mathrm{\mu s}$). The individual contributions were estimated by calculating accidental coincidences of measured rates (a, b) and by extrapolating the delayed coincidence signatures into the MiDM signal region (c, d), taking into account all requirements on the energies of the signals, timing and interaction sequence (NR $\to$ ER). The contributions (a)-(c) turn out to be negligible and the background expectation for the total exposure is $(0.17\pm0.11)$\,events in the signal region, given by the Bi-Po background. In principle, these backgrounds could be further lowered by requiring a large spatial distance between the two interaction vertices, however, due to the exponential decay time spectrum of the WIMP de-excitation, such condition would significantly affect the detection efficiency.

%%%%%%%%%%%%%%%%%%%%%%%%%%%%%%%%%%%%%%%%%%%%%%%%%%%%%%%%%%%%%%%%%%%%%%%%%%%%%%
\section{\label{sec:efficiency}Detection efficiency}

The main signal loss in this analysis comes from to the finite size of the XENON100 detector which limits the chance to detect the $\gamma$-ray from the WIMP de-excitation. This detection efficiency is determined with a Monte Carlo simulation based on the approach used in~\cite{yavin}. It simulates the position, time and energy of the NR and ER signals from MiDM interactions inside the XENON100 TPC.  At first, values for the WIMP velocity $\mathbf{v}$ and the recoil energy $E_R$ are randomly generated according to the differential rate d$R$/d$E_R$. To speed up the simulations, the differential cross section d$\sigma$/d$E_R$ is approximated by using the dipole charge term d$\sigma_\mathrm{DZ}$/d$E_R$ only, see equation~(\ref{eq::dz}), which contributes about 80\% to the total cross section for a Xe target~\cite{Lin:directional_signal}. We verified that this approximation affects the outcome of the analysis only at the $<$2\,\% level and will always lead to more conservative results. 

\begin{figure}[]
\centering
\begin{minipage}[t]{.5\textwidth}
  \centering
  \includegraphics[width=.95\linewidth]{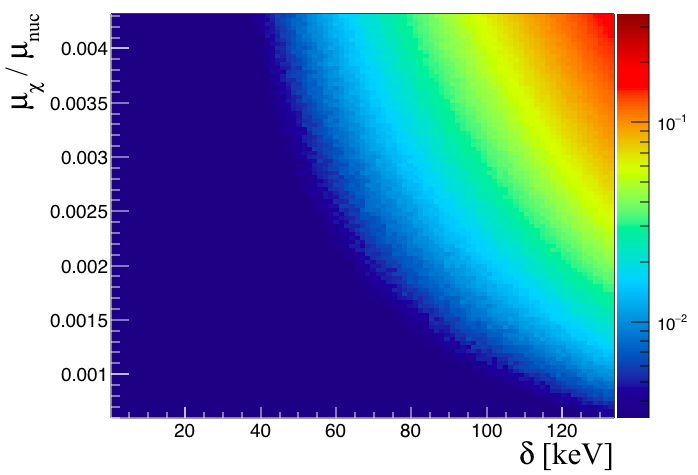}
\end{minipage}%
\begin{minipage}[t]{.5\textwidth}
  \centering
  \includegraphics[width=.95\linewidth]{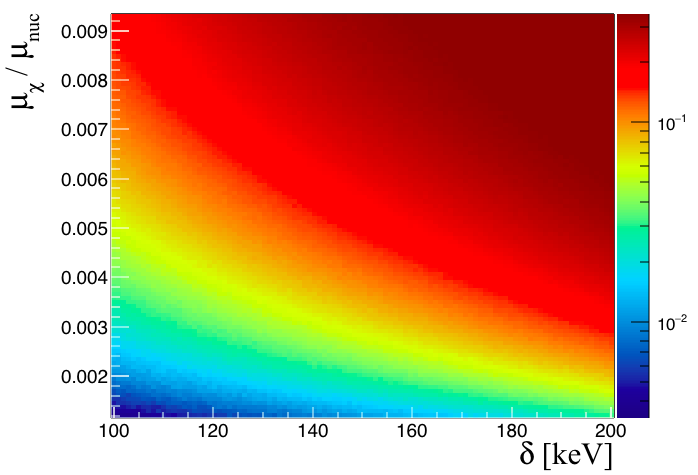}
\end{minipage}
  \caption{Simulated efficiency (given by the color scale) for detecting both the NR and the de-excitation ER signal, inside the 48\,kg fiducial volume of XENON100 for the considered range of mass splittings~$\delta$ and WIMP magnetic moments $\mu_\chi$ (in units of the nuclear magneton $\mu_{\mathrm{nuc}}$). It is shown for the two benchmark cases, corresponding to WIMP masses of $m_\chi=58.0$\,GeV/c$^2$ (left) and $m_\chi=122.7$\,GeV/c$^2$ (right)~\cite{Barello:2014}. For $m_\chi=58.0$\,GeV/c$^2$, the efficiency is significantly smaller, mainly due to the smaller $\delta$ and  $\mu_\chi$ which leads to a longer lifetime $\tau=\pi/(\delta^3\mu_\chi^2)$, and thus to a longer mean path length until the WIMP de-excites.}
\label{fig:efficiency}
\end{figure}

The velocity $\mathbf{v}^\prime$ of the WIMP after scattering can be calculated from the scattering kinematics, taking into account the diurnal rotation of the detector with respect to the WIMP wind. The positions $\mathbf{x}$ of NR events are uniformly distributed inside the XENON100 TPC as expected from the small WIMP interaction cross section. The vertex $\mathbf{x}^\prime$ of the de-excitation is calculated via
\begin{equation}
\mathbf{x}^\prime = \mathbf{x}+\Delta t \cdot \mathbf{v}^\prime,
\end{equation}
where $\Delta t$ is sampled from an exponential distribution using the lifetime $\tau$ of the excited WIMP. $\mathbf{x}^\prime$ is taken to be the position of the ER. The range of the $\mathcal{O}$(100\,keV) photon emitted during the de-excitation is only about 2\,mm and neglected here.

For both recoil events, the time of the S2 signals is calculated according to their time difference $\Delta t$ and their z-position inside the TPC using an electron drift velocity of 1.73\,mm/$\mu$s~\cite{XENON100:instrument}. Finally, the detection efficiency is calculated as
\begin{equation}
\epsilon=\frac{N_\mathrm{det}}{N_\mathrm{all}}\,.
\end{equation}
$N_\mathrm{det}$ is the number of events with both signals (NR and ER) being located inside the 48\,kg fiducial target and which also fulfill the additional timing cuts employed in the analysis of the data: the time difference between the S1 signals is $>$50\,ns and the time difference between the S2 signals is $>$3.5\,$\mu$s; the first S2 signal comes after the second S1 signal.
The number $N_\mathrm{all}$ denotes all events where the nuclear recoil is located inside the fiducial volume. The resulting efficiencies for the two benchmark cases motivated by the best-fits to DAMA/LIBRA are shown in Figure~\ref{fig:efficiency}. The largest source of systematic uncertainty is the approximation of the cross section, however, its impact on the result is well below statistical fluctuations and therefore negligible.

%%%%%%%%%%%%%%%%%%%%%%%%%%%%%%%%%%%%%%%%%%%%%%%%%%%%%%%%%%%%%%%%%%%%%%%%%%%%%%
\section{\label{sec:results}Result and conclusions}

\begin{figure}[]
\centering
\begin{minipage}[t]{.5\textwidth}
  \centering
  \includegraphics[width=1\linewidth]{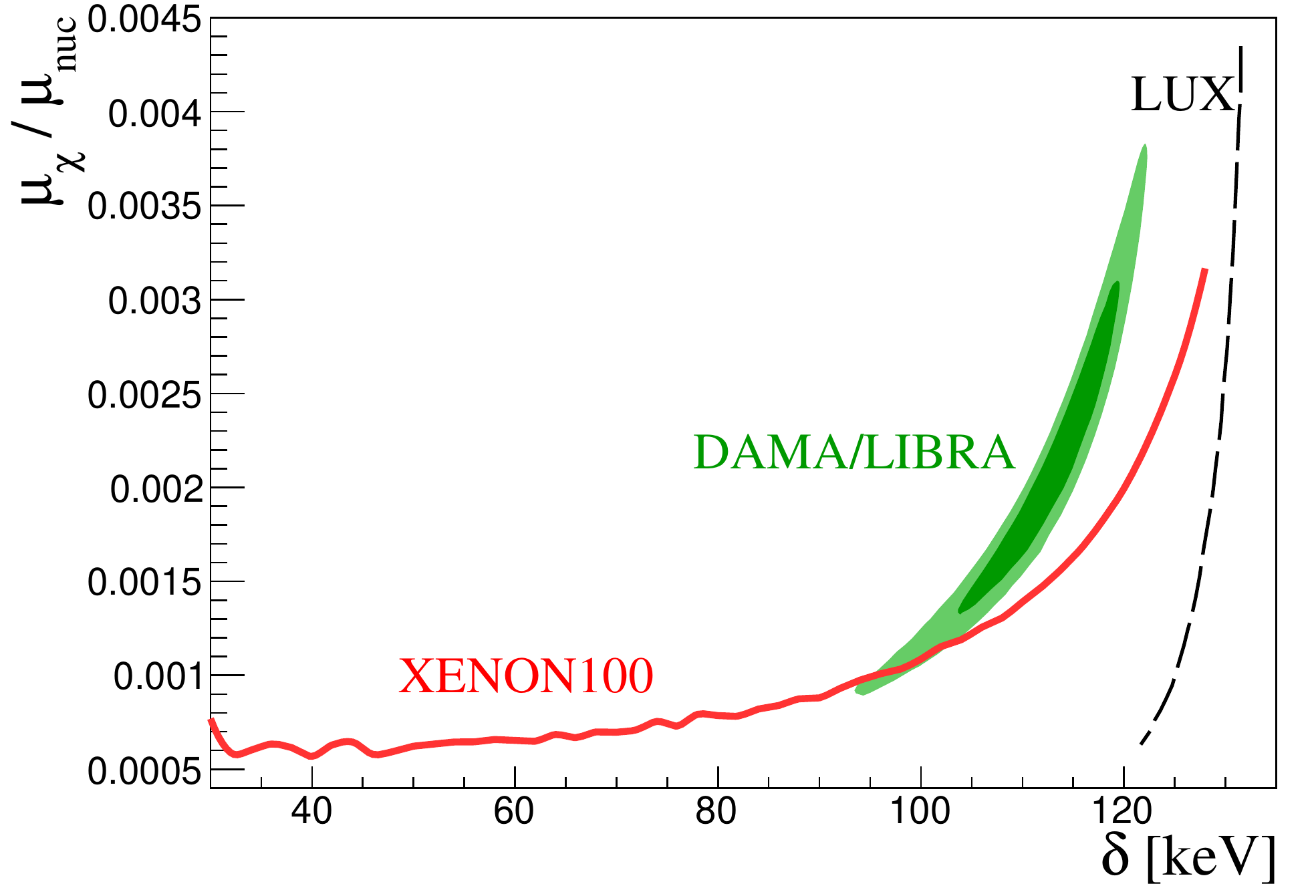}
\end{minipage}%
\begin{minipage}[t]{.5\textwidth}
  \centering
  \includegraphics[width=1\linewidth]{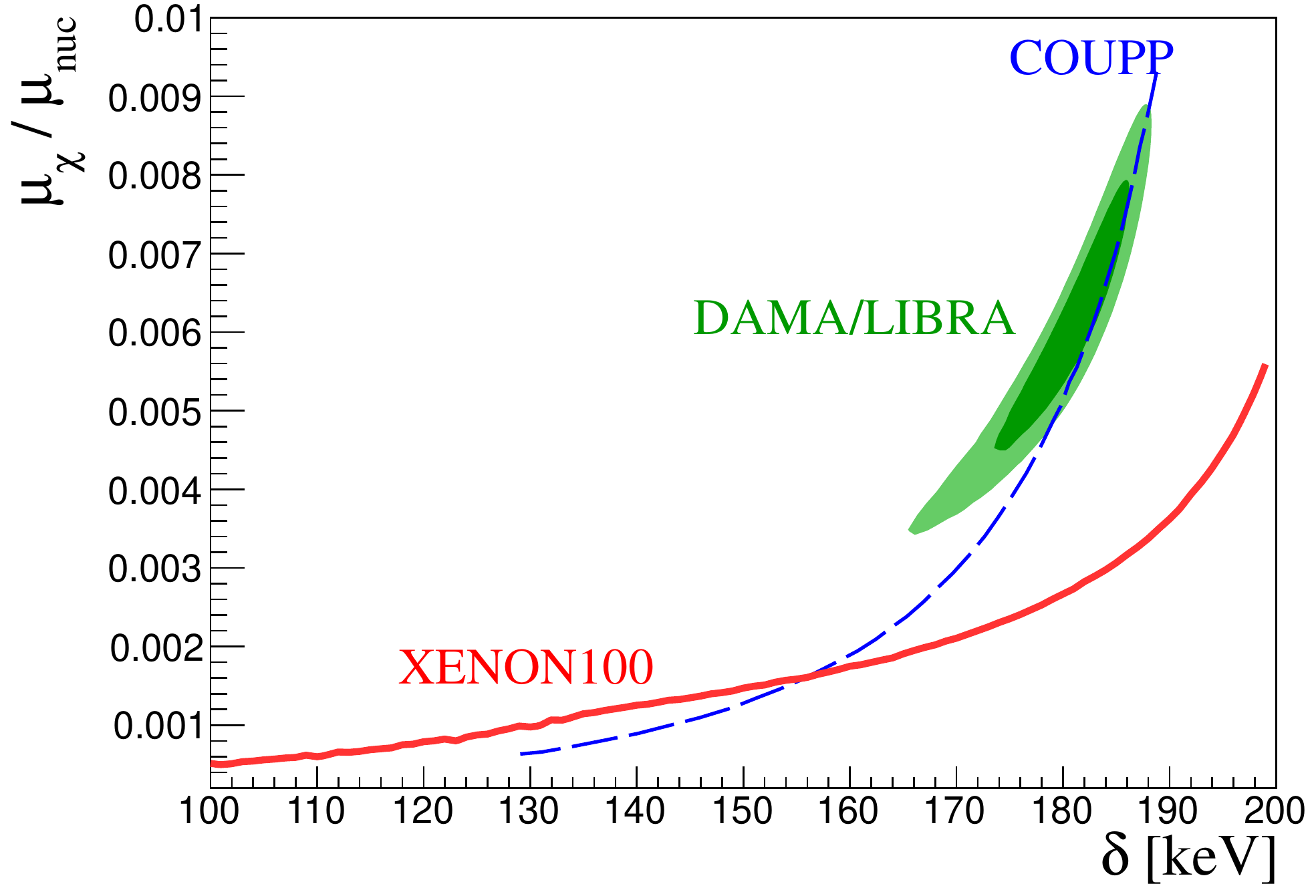}
\end{minipage}
\caption{The exclusion limit (at 90\% confidence level, CL) on MiDM interactions from the run~II of XENON100 is shown by the red curve for WIMP masses of $m_\chi=58.0$\,GeV/c$^2$ (left) and $m_\chi=122.7$\,GeV/c$^2$ (right). Also shown are the 68\% (dark green) and 95\% (light green) CL regions of the best fit to the DAMA/LIBRA modulation signal~\cite{Barello:2014}. Limits calculated in~\cite{Barello:2014} using results from LUX and COUPP are shown for comparison (dashed lines). For a WIMP mass of $m_\chi=122.7$\,GeV/c$^2$ (right), the XENON100 result based on the search for two subsequent signals is superior to the previous result above $\delta\approx 155$\,keV and rules out the entire best-fit region.}
\label{fig:limit_plots}
\end{figure}

\begin{figure}
  \begin{minipage}[c]{0.6\textwidth}
    \includegraphics[width=0.95\textwidth]{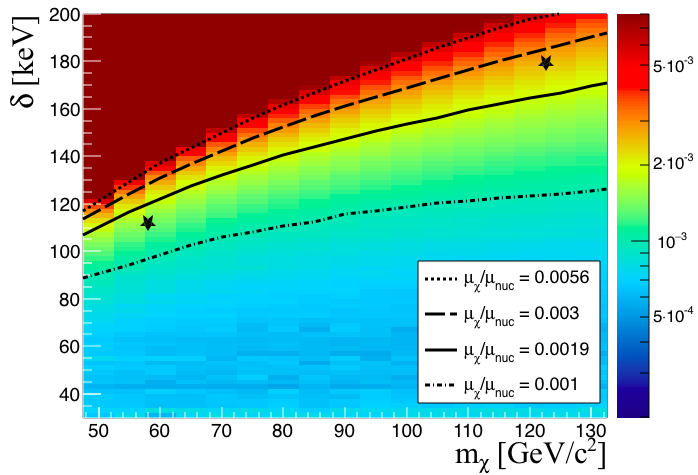}
  \end{minipage}\hfill
  \begin{minipage}[c]{0.4\textwidth}
  \vspace{-2.4cm}
    \caption{The exclusion limit (at 90\% CL) on MiDM interactions for a wide range of masses~$m_\chi$ and mass-splittings $\delta$. The excluded magnetic moment~$\mu_{\chi}/\mu_{nuc}$ is given by the color scale. The four exemplary lines mark contours of equal $\mu_{\chi}/\mu_{nuc}$. The stars indicate the two benchmark cases presented in Figure~\ref{fig:limit_plots}.}
    \label{fig:result_2d}
  \end{minipage}
\end{figure}

After applying the data selection cuts described in Section~\ref{sec:analysis}, no MiDM candidate event has been found in the XENON100 science run~II dataset with a total exposure of 10.8\,ton\,$\times$\,days, see Figure~\ref{fig:ROI} (left). We calculate an upper limit on the interaction strength using the maximum gap method~\cite{Yellin:MaxGap}. Figure~\ref{fig:limit_plots} shows the resulting limits for the two benchmark cases, corresponding to $m=58.0$\,GeV/c$^2$ and $m=122.7$\,GeV/c$^2$, together with the 68\,\% and 95\,\% confidence level regions for DAMA/LIBRA taken from~\cite{Barello:2014}. 
The two best-fit points to the DAMA/LIBRA modulation signal are excluded at 3.3\,$\sigma$ (1: $m_\chi$\,=\,58.0\,GeV/$c^2$, $\mu_\chi$\,=\,0.0019 \,$\mu_{\mathrm{nuc}}$, $\delta$\,=\,$111.7$\,keV) and 9.3\,$\sigma$ (2: $m_\chi$\,=\,122.7\,GeV/$c^2$, $\mu_\chi$\,= 0.0056\,$ \mu_{\mathrm{nuc}}$, $\delta$\,=\,179.3\,keV). 
The analysis relies on the detection of both interactions (NR and ER de-excitation), an approach which has not yet been pursued in a dark matter analysis so far. Therefore, the sensitivity towards lower mass splittings is not competitive to previous results~\cite{Barello:2014}, where only the NR interaction is taken into account. 
However, at higher $\delta$ and thus shorter lifetimes of the excited WIMP, a significant improvement of the limits on the MiDM interaction strength is achieved. 
This can also be seen in Figure~\ref{fig:result_2d}, where the 90\,\% confidence level exclusion limits are presented for a wide range of parameters $(m_\chi,\delta)$.
While the DAMA/LIBRA best-fit region has already been ruled out for benchmark case~1 ($Q_\text{I}=0.09$, $m_\chi=58.0$\,GeV/$c^2$) in~\cite{Barello:2014} by using data from LUX~\cite{LUX:2013}, our new analysis now also excludes the DAMA/LIBRA modulation signal being due to MiDM interactions assuming the newer quenching factor $Q_\text{I}=0.04$ (benchmark case~2), corresponding to $m_\chi=122.7$\,GeV/c$^2$, and covers previously unexplored parameter space above $\delta\approx 155$\,keV. The exclusion limits on MiDM interactions for arbitrary combinations of $m_\chi$ and $\delta$, i.e., without any reference to the DAMA/LIBRA experiment, are also presented for the first time.

The sensitivity of this type of analysis will be greatly improved for current ton-scale (e.g., XENON1T~\cite{ref:xe1t,ref:xe1tSR0}) and future multi-ton dual-phase LXe TPCs (e.g., XENONnT~\cite{ref:xe1t}, LZ~\cite{ref::lz} and DARWIN~\cite{ref::darwin}). This is not only due to the increased target mass, but also thanks to the higher probability of detecting the de-excitation inside the larger active volume. The specific MiDM signature of two~S1 followed by two~S2 signals differs significantly from the most common backgrounds and leads to a very low background expectation while exploiting a large fraction of the target mass.

\acknowledgments 
We thank Itay Yavin for providing us the code of his efficiency simulation and the useful discussion and we thank Spencer Chang and Chris Newby for providing us their code to calculate the differential event rate. 
We gratefully acknowledge support from the National Science Foundation, Swiss National Science Foundation, Deutsche Forschungsgemeinschaft, Max Planck Gesellschaft, German Ministry for Education and Research, Netherlands Organisation for Scientific Research (NWO), Weizmann Institute of Science, I-CORE, Initial Training Network Invisibles (Marie Curie Actions, PITNGA-2011-289442), Fundacao para a Ciencia e a Tecnologia, Region des Pays de la Loire, Knut and Alice Wallenberg Foundation, Kavli Foundation, and Istituto Nazionale di Fisica Nucleare. We are grateful to Laboratori Nazionali del Gran Sasso for hosting and supporting the XENON project.

\end{document}